\documentclass[pdflatex,sn-nature]{sn-jnl}

\usepackage{graphicx}%
\usepackage{tabularx} %
\usepackage{multirow}%
\usepackage{amsmath,amssymb,amsfonts}%
\usepackage{amsthm}%
\usepackage{mathrsfs}%
\usepackage[title]{appendix}%
\usepackage{xcolor}%

\usepackage{textcomp}%
\usepackage{manyfoot}%
\usepackage{booktabs}%
\usepackage{algorithm}%
\usepackage{algorithmicx}%
\usepackage{algpseudocode}%
\usepackage{listings}%
\usepackage[center]{caption}
\usepackage{pdflscape}
\usepackage{float}
\usepackage{geometry}
\usepackage{tcolorbox}
\usepackage{placeins}
\theoremstyle{thmstyleone}%
\theoremstyle{thmstyletwo}%

\theoremstyle{thmstylethree}%

\begin{document}

\title[Article Title]{Auditing Sex/Gender Disparities in Emergency Triage with LLM-based Paired Comparisons}


\author[1,2,*]{\fnm{Ariel} \sur{Guerra-Adames}}\email{ariel.guerra-adames@u-bordeaux.fr}

\author[2]{\fnm{Marta} \sur{Avalos-Fernandez}}

\author[1]{\fnm{Océane} \sur{Dorémus}}

\author[3,4,5]{\fnm{Leo Anthony} \sur{Celi}}

\author[1,6]{\fnm{Cédric} \sur{Gil-Jardiné}}

\author[1]{\fnm{Emmanuel} \sur{Lagarde}}

\affil[1]{\orgdiv{AHeaD Team, Bordeaux Population Health Research Center}, \orgname{Université de Bordeaux}, \orgaddress{\city{Bordeaux}, \postcode{F-33000}, \country{France}}}

\affil[2]{\orgdiv{SISTM Team, Bordeaux Population Health Research Center}, \orgname{Université de Bordeaux}, \orgaddress{ \city{Bordeaux}, \postcode{F-33000}, \country{France}}}

\affil[3]{\orgdiv{Laboratory for Computational Physiology}, \orgname{Massachusetts Institute of Technology}, \orgaddress{ \city{Cambridge}, \country{USA}}}

\affil[4]{\orgdiv{Division of Pulmonary, Critical Care and Sleep Medicine}, \orgname{Beth Israel Deaconess Medical Center}, \orgaddress{ \city{Boston}, \country{USA}}}

\affil[5]{\orgdiv{Department of Biostatistics}, \orgname{Harvard T.H. Chan School of Public Health}, \orgaddress{ \city{Boston}, \country{USA}}}

\affil[6]{\orgdiv{Emergency Department}, \orgname{CHU de Bordeaux}, \orgaddress{\city{Bordeaux}, \postcode{F-33000}, \country{France}}}


\abstract{We present a domain-agnostic paired-comparison approach that uses Large Language Models (LLMs) to quantify sex/gender-related asymmetries in documented clinical decision-making. The method trains an LLM to emulate observed decisions, then evaluates sex-swapped pairs in which only sex is flipped, holding documented clinical content constant. We apply it to emergency triage, analyzing more than 140{,}000 Bordeaux University Hospital (France) admissions and testing methodological portability on MIMIC-IV, spanning a different language, population, and healthcare system. Fine-tuning Mistral NeMo 12B for triage prediction and using Mistral Small 24B for pair generation, we find otherwise identical presentations were more likely to receive a lower-severity predicted score as female than male: 1.1\% (95\% CI 0.9--1.3) in the French cohort, 2.2\% (1.7--2.7) in MIMIC-IV. Predictions are sensitive to both tabular and textual sex markers, with the asymmetry emerging primarily in the combined bimodal setting. A model retrained on sex-neutralized inputs eliminated the between-sex prediction gap, indicating the asymmetry is mediated by explicit sex markers. Patterns vary with nurse-patient sex concordance, suggesting the model captures stable features of the recorded data rather than random artifacts. These effects are small and documentation-level. We therefore present this as a methodological feasibility study: LLMs can serve as scalable probes of documented decisions, generating hypotheses rather than establishing bedside clinician behavior or clinically meaningful undertriage, which would require clinician-anchored validation. Beyond emergency care, the approach supports bias audits in other domains.}

\maketitle

\section*{Introduction}\label{introduction}

Emergency triage represents one of the most critical decision points in healthcare delivery, where trained professionals must rapidly assess patient acuity and allocate limited resources under significant time pressure. Modern emergency departments employ standardized triage systems, such as the 5-level Emergency Severity Index (ESI) in the United States or the French Emergency Nurses Classification in Hospitals (FRENCH), to stratify patients based on clinical urgency \cite{zachariasse2019performance}. These systems aim to ensure that patients with the most critical conditions receive immediate care, with triage levels ranging from 1 (requiring immediate resuscitation) to 5 (non-urgent). The stakes of these decisions are profound: undertriage can delay critical interventions and increase mortality risk, while overtriage can overwhelm emergency resources and compromise care for other patients \cite{hinson2019triage}.

Despite standardization efforts, mounting evidence suggests that emergency triage decisions are susceptible to systematic biases related to patient demographics. Multiple studies have documented disparities in triage assignment based on sex (biological and physiological characteristics of males and females) or gender (socially constructed characteristics) \cite{coisy2023triageappearance, han2023suicidebias, banco2022chestpain, onal2022throughput, lopez2021poisoning, preciado2021acs, vigil2016patient, chen2008gender, croskerry2001emergency, arslanian-engoren2000triage}, race \cite{schrader2013racial, lopez2010racial, lin2022prioritization, peitzman2023subjective, morel2019france, wu2023emsStroke}, age \cite{platts2010accuracy}, and socioeconomic status \cite{kangovi2013understanding, turner2022inequality, verma2023emsDelays}. Gender bias, in particular, has been observed across diverse clinical presentations. Women presenting with acute coronary syndrome are more likely to receive lower triage acuity scores compared to men \cite{mnatzaganian2016sex}. Similar patterns have been documented for stroke and abdominal pain \cite{amy2019sex}, where women experience longer wait times and higher rates of undertriage. These disparities persist even after controlling for clinical factors, suggesting the influence of implicit biases in clinical decision-making \cite{fitzgerald2017implicit}.

It is important to emphasize that, to date, there is no systematic ``gold standard'' for evaluating triage decisions that would allow one to determine, for each patient, whether they have been over- or under-triaged. While several studies illustrate the usefulness of audit in improving triage quality \cite{ouellet2025triage, zaboli2023audit} and provide examples of how such audits can be conducted \cite{pilleron2024moral}, they do not establish universal reference standards. Patient outcomes may offer some indication of the appropriateness of a triage decision and the association with patient's sensitive attributes \cite{liu2024prognostic}, but they cannot account for the contextual information available to the triage nurse at the time of assessment. Throughout this paper, we use the term ``bias'' to refer specifically to \textit{differential treatment of clinically identical presentations for which there is no clinical justification for different treatment}. In the context of emergency triage, bias manifests when two patients presenting with the same clinical picture (i.e., identical symptoms, vital signs, medical history, and urgency indicators) receive different triage scores based solely on non-clinical attributes such as gender. This operational definition focuses on unjustified disparities rather than all outcome differences, acknowledging that some differences in treatment may be clinically appropriate.

The emergence of artificial intelligence, particularly Large Language Models (LLMs), offers unprecedented opportunities to both understand and address these biases. Crucially, while LLMs have demonstrated remarkable capabilities in clinical tasks \cite{thirunavukarasu2023large, singhal2023large, patel2023chatgpt, shen2021multimodal}, they also notoriously inherit and reproduce many unintended human behaviors, including the systematic biases present in their training data \cite{mahajan2025cognitive, yang2024unmasking, omiye2023large}. This tendency, traditionally viewed as a significant limitation, presents a unique opportunity: \textbf{LLMs can serve as ``bias mirrors,''} providing a scalable tool for auditing human decisions. Rather than deploying LLMs to replace human decision-makers, we propose leveraging their ability to faithfully reproduce human behavior, including its biases, as an analytical tool to detect and quantify these disparities in clinical practice. Their ability to process both structured and unstructured clinical data makes them particularly well-suited for emergency triage, where decisions integrate vital signs, chief complaints, and clinical narratives. By training an LLM to approximate existing human triage decisions as recorded in clinical documentation, we aim to obtain a model that reflects patterns of clinical reasoning and potential bias embedded in current documented practice.

\begin{figure}[H]
\centering
\caption*{\textbf{Overview of the paired-comparison framework for gender bias detection in emergency triage.}}
\includegraphics[width=0.9\linewidth]{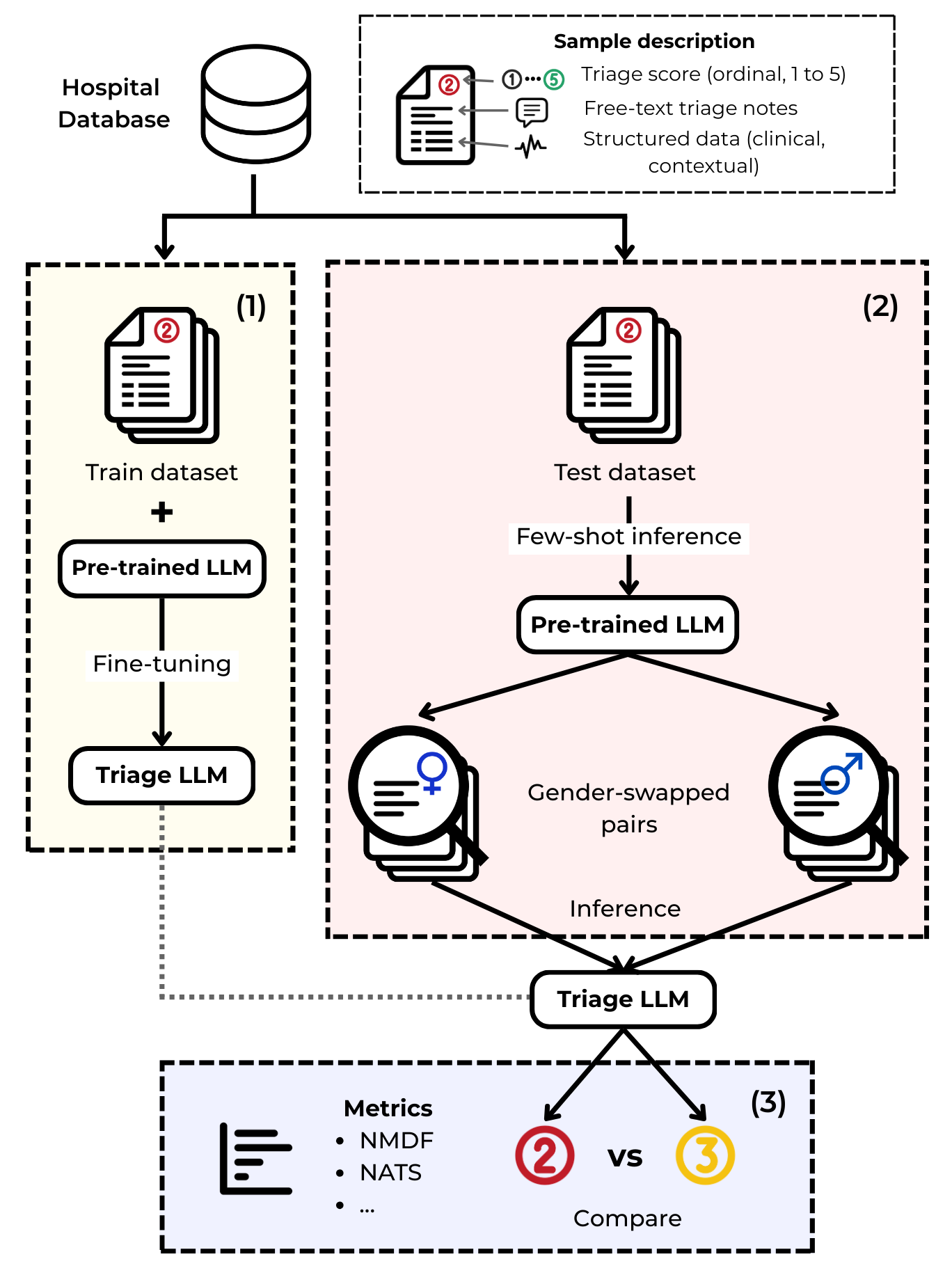}
\caption{The workflow consists of (1) LLM fine-tuning for emergency triage prediction, (2) generation of gender-swapped pairs by flipping sex/gender while preserving clinical content, and (3) comparison of triage predictions on matched pairs.}
\label{fig:workflow}
\end{figure}

In a previous study, we developed a paired-comparison framework based on LLMs to detect bias in emergency triage decisions \cite{guerra2025uncovering}. As illustrated in Figure~\ref{fig:workflow}, this framework consists of three interconnected components: (1) a fine-tuned LLM-based triage model that replicates human triage decision-making, (2) an LLM-based gender-swapped sample generator that systematically modifies sensitive attributes while preserving clinical content, and (3) a set of triage-specific inequality metrics for bias quantification. The framework takes as input a dataset of patient admissions from an emergency department (ED), where each record contains an ordinal triage score assigned by trained clinical personnel, free-text triage notes documenting the clinical assessment and circumstances, and structured variables capturing vital parameters and contextual factors. By comparing triage predictions between original and gender-swapped presentations, the framework is designed to detect unjustified differential treatment by comparing predictions on matched pairs that differ only in gender presentation.

That initial work introduced the core methodological concept but left several key questions unanswered: How can systematic disparities be rigorously quantified through appropriate bias metrics? How can we isolate the association between patient gender and triage outcomes while controlling for clinically relevant variables? What is the potential impact on patient care at scale? And critically, are the observed patterns reproducible across different datasets, languages, countries, and healthcare systems? The present study addresses these questions by developing appropriate metrics and applying the framework to two distinct datasets: a French dataset from the Bordeaux University Hospital protected under European (GDPR) and national (CNIL) regulations, and a subset of the semi-publicly accessible MIMIC-IV database from the United States. Throughout, it should be emphasized that the study measures the sensitivity of model predictions to sex/gender markers present in documented triage records. It does not directly observe or measure bedside clinician behavior; rather, it uses documentation-conditioned model behavior as a hypothesis-generating audit of possible sex/gender-related asymmetries in recorded triage decisions. Consistent with this scope, we frame the present work primarily as a methodological feasibility study: its principal contribution is to establish that a fine-tuned LLM can be used as a scalable probe of documented decision data, with the size and direction of any detected sex/gender-marker effect reported as an empirical, hypothesis-generating finding.

\section*{Results} \label{sec:results}

\noindent\textbf{Triage prediction task.}
To select the most suitable architecture for our bias analysis, we evaluated a range of models on the emergency triage task using the full test partition of the Bordeaux University Hospital ED dataset ($n = 72{,}444$). Performance was primarily assessed using quadratically weighted Cohen's kappa ($\kappa_w$), a standard measure for triage agreement among human experts \cite{suamchaiyaphum2024accuracy, pourasghar2014kappa}, alongside precision, recall, and macro-averaged F1-score.

We compared classical machine learning and deep-learning architectures, from feature-based baselines to medium- and large-scale LLMs. Model descriptions are provided in Methods. All deep-learning models were fine-tuned until training loss stabilized (typically 3--5 epochs). Additional training details are provided in the Methods.

\begin{figure}[H]
\centering
\caption*{\textbf{Model triage performance ($\kappa$).}}
\includegraphics[width=0.9\linewidth]{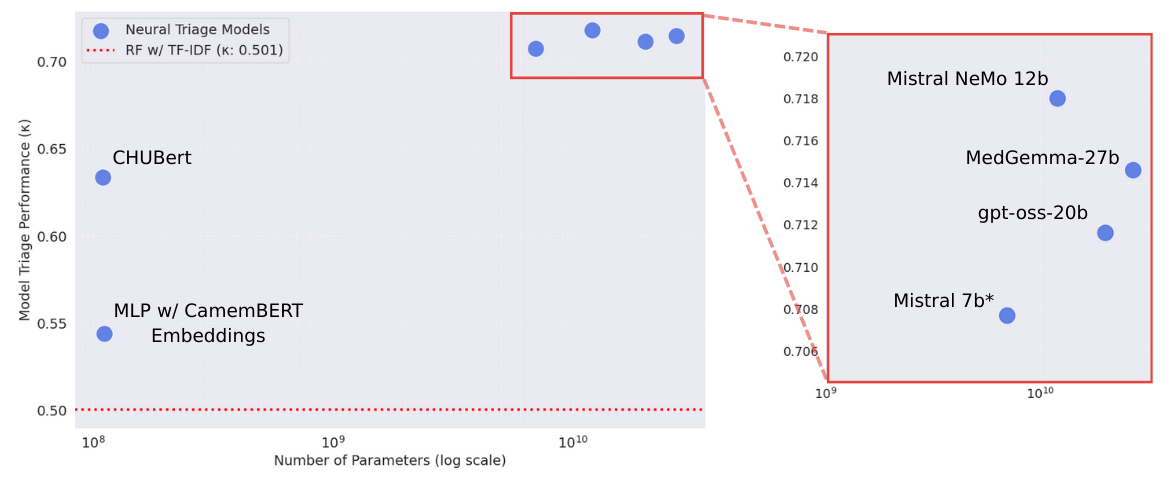}
\caption{Performance of triage models on the Bordeaux University Hospital ED test set, evaluated with quadratically weighted Cohen's $\kappa$. The left panel shows all evaluated architectures ordered by parameter size; the red dashed horizontal line indicates the RF/TF-IDF baseline. The right panel provides a zoomed view of the large-LLM region ($>10^9$ parameters) to resolve performance differences among models with similar $\kappa_w$.}
\label{fig:model_performance}
\end{figure}

Figure~\ref{fig:model_performance} summarizes the relationship between model size and $\kappa_w$; Table~\ref{tab:model_performance} reports the full set of metrics.

Across the evaluated architectures, we observed a clear performance gradient from classical approaches to larger LLMs. The RF/TF-IDF baseline performed the weakest ($\kappa_w=0.5006$), while transformer-based models, even at moderate parameter scales such as CHUBert ($3.3 \times 10^{8}$), yielded substantial gains ($\kappa_w=0.6339$). Among medium-to-large LLMs, MistralNeMo 12B achieved the highest $\kappa_w$ (0.7180), with MedGemma 27B a close second in $\kappa_w$ (0.7146) despite being more than twice as large. The best precision (0.6692) was obtained by Mistral 7B; these results originate from our prior study \cite{guerra2025uncovering} and are included here for reference. Interestingly, scaling beyond 12B parameters did not yield substantial improvements in $\kappa_w$, suggesting diminishing returns for this task at higher model capacities. This performance is noteworthy given that published studies report human expert inter-rater agreement of $\kappa = 0.61-0.81$ for structured triage systems such as ESI and MTS \cite{hernandez2015esi, olofsson2009mts, ganjali2020esiaccuracy}. Our model's $\kappa_w = 0.718$ falls within this range of human expert agreement, indicating that the model reproduces documented triage labels with reliability comparable to reported human inter-rater agreement, while noting that this reflects agreement with recorded labels rather than the full bedside triage process. Based on these results, we selected Mistral NeMo 12B as the triage prediction model for all subsequent bias analyses due to its size-peformance balance.

\begin{table}[h]
\centering
\begin{tabularx}{\textwidth}{l>{\centering\arraybackslash}X>{\centering\arraybackslash}X>{\centering\arraybackslash}X>{\centering\arraybackslash}X>{\centering\arraybackslash}X>{\centering\arraybackslash}X>{\centering\arraybackslash}X}
\hline
\textbf{Model} & \textbf{Size} & $\boldsymbol{\kappa_w}$ & \textbf{Precision} & \textbf{Recall} & \textbf{F1 Score} & \textbf{WOA} \\
\hline
RF w/ TF-IDF & -- & 0.5006 & 0.6103 & 0.5699 & 0.5346 & 0.9776 \\
MLP w/ BERT Embeddings & $3.3\times 10^8$ & 0.5439 & 0.5716 & 0.5733 & 0.5559 & 0.9736 \\
CHUBert & $3.3\times 10^8$ & 0.6339 & 0.6133 & 0.6133 & 0.6035 & 0.9849 \\
\textbf{Mistral NeMo 12B} & $1.2\times 10^{10}$ & \textbf{0.7180} & 0.6640 & 0.6656 & 0.6615 & \textbf{0.9892} \\
gpt-oss-20b & $2.1\times 10^{10}$ & 0.7116 & 0.6667 & \textbf{0.6677} & \textbf{0.6648} & \textbf{0.9892} \\
MedGemma 27B & $2.7\times 10^{10}$ & 0.7146 & 0.6585 & 0.6582 & 0.6580 & 0.9851 \\
\hline
Mistral 7B~\cite{guerra2025uncovering} & $7\times 10^9$ & 0.7077 & \textbf{0.6692} & 0.6676 & 0.6633 & -- \\
\hline
\end{tabularx}
\caption{Performance metrics on the test set. For each metric, the best performance is highlighted in bold. The Mistral 7B row reports results from our previous work using similar data~\cite{guerra2025uncovering}. WOA = Within-One Accuracy (predictions within $\pm$1 of true triage score).}
\label{tab:model_performance}
\end{table}

\vspace{0.2cm}
\noindent\textbf{Gender-Swapped Pairs Generation Task.}
We manually assessed the quality of the Mistral Small 24B-generated gender-swapped pairs by having two independent annotators review a stratified random sample of 500 transformations from the Bordeaux University Hospital ED dataset and 250 from the MIMIC-IV dataset. The two annotators were (1) a doctoral researcher in digital public health with training in clinical data analysis and medical terminology (A.G.A.), and (2) an emergency medicine physician at the Bordeaux University Hospital (C.G.J.). Each transformation was classified into one of three categories: (1) correct and complete transformation, where all gender references were accurately modified without changing the rest of the content in the clinical notes, (2) incomplete transformation, where the main gender reference was changed but some secondary gendered language or pronouns were missed, or where medical details were altered and (3) failed transformations, where the transformation changed clinical information other than gender references or added hallucinated information. For the Bordeaux University Hospital ED subset (n=500), annotators agreed on 494 samples and disagreed on 6, which were labeled either as correct or incomplete by annotators. For the MIMIC-IV subset (n=250), annotators agreed on 248 samples and disagreed on 2, which were also labeled either as correct or incomplete. No samples were labeled as failed in either dataset.

Semantic invariance diagnostics further confirmed that the transformations preserved clinical content: on MIMIC-IV, mean embedding cosine similarity between original and gender-swapped notes was $0.9996$; on the Bordeaux University Hospital ED dataset, mean cosine similarity was $0.9694$, with the lower value attributable to obligatory French grammatical gender agreement cascading across determiners, adjectives, and participles. Full results are reported in Supplementary Note~5, Supplementary Table~11, and Supplementary Figure~4. Together with the manual review above, these checks support the linguistic and semantic stability of the counterfactual rewriting procedure, that is, that the edits change sex/gender references while leaving the documented clinical content intact. They are not a substitute for clinical re-triage, chart review, or outcome-based validation, which would be required to establish the clinical significance of the resulting prediction shifts.

\vspace{0.2cm}
\noindent\textbf{Detection of gender biases.}
We conducted the bias analysis using Mistral NeMo 12B fine-tuned for triage prediction on the full test partition of the Bordeaux University Hospital ED dataset (40,082 originally male and 32,362 originally female records), under three transformation conditions: full, text-isolated, and tabular-isolated (Table~\ref{tab:bias_metrics_side_by_side}). For a methodological portability test, we repeat the full-transformation analysis on the MIMIC-IV test partition (21,342 originally male and 16,874 originally female patient records). Metrics that are proportions (e.g., DTS, NATS) are reported as percentages. Throughout, we describe changes in terms of higher vs.\ lower predicted triage scores, without normative terminology.

We foreground the Directional Triage Skew (DTS), which, within each transformation direction, measures the net balance between increases and decreases in the predicted triage score. In the full setting on the Bordeaux University Hospital ED dataset, we obtained $ \mathrm{DTS}_{M\mid F}=-0.8\%\;[-1.1,\,-0.5] $ and $ \mathrm{DTS}_{F\mid M}=1.3\%\;[1.0,\,1.6] $. When transforming samples from female→male, decreases in the predicted score occurred more often than increases by 0.8 percentage points. When transforming samples from male→female, increases occurred more often than decreases by 1.3 points.

These two skews are estimated on disjoint cohorts and each measures the same female-versus-male contrast from one side. We therefore summarize them as a single size-weighted pooled per-pair rate rather than as their sum; the derivation and per-cohort breakdown are given in Supplementary Note~6 and Supplementary Table~12. The pooled rate was $1.09\%\;[0.89,\,1.29]$ (bootstrap 95\% CI): for otherwise identical presentations, the female variant was about $1.1$ percentage points more likely than the male variant to receive a higher (less severe) predicted triage score.

The Pairwise Disagreement Rate (PDR) was $8.4\%\;[8.2,\,8.6]$, indicating that the transformation changed the assigned class in a non-trivial minority of cases.

Modality-isolated analyses show complementary patterns. Under text-isolated transformations, both skews were substantially negative, $ \mathrm{DTS}_{F\mid M}=-10.3\%\;[-10.8,\,-9.9] $ and $ \mathrm{DTS}_{M\mid F}=-12.3\%\;[-12.8,\,-11.9] $, meaning that editing only textual gender cues tends to produce more decreases than increases in the predicted score in both directions. In contrast, under tabular-isolated transformations, both skews were positive and similar in magnitude, $ \mathrm{DTS}_{F\mid M}=+3.8\%\;[3.2,\,4.3] $ and $ \mathrm{DTS}_{M\mid F}=+3.8\%\;[3.2,\,4.4] $, meaning that flipping only the tabular sex field tends to produce more increases than decreases in the predicted score in both directions. These modality-specific inversions suggest that explicit tabular sex markers and implicit textual gender cues influence predictions in opposite directions, with text edits prompting more frequent decreases and tabular flips prompting more frequent increases in the predicted triage score.

The Net Mean Difference (NMD; formally defined in Supplementary Note~1) measures the average signed difference in predicted triage scores between the female and male versions of the same clinical content; a positive value indicates that the female variant receives a higher (less severe) predicted score on average. In the Bordeaux University Hospital ED dataset, $ \mathrm{NMD}=0.011\;[0.009,\,0.013] $ and, in the tabular-isolated setting, $ \mathrm{NMD}=0.006\;[0.001,\,0.010] $. Because NMD is a mean difference on the ordinal triage scale (not a proportion), absolute values are small; the positive sign indicates that, on average, the female variant receives a slightly higher (less severe) triage score than the male variant for the same clinical content. We emphasize that, given the very large sample sizes, intervals can exclude zero even when the underlying magnitude is small: an NMD of roughly $0.01$ point on the five-level scale is statistically distinguishable from zero but is unlikely to be clinically meaningful for any individual patient, and should be read as a population-level documentation-level signal rather than evidence of individual harm.

Under text isolation, $ \mathrm{NATS}(+)=-0.8\%\;[-1.0,\,-0.5] $ and $ \mathrm{NATS}(-)=1.2\%\;[0.7,\,1.7] $ meaning that higher-score changes are more frequent for male$\to$female transformations than female$\to$male, while lower-score changes are more frequent for female$\to$male than male$\to$female transformations; under tabular isolation, $ \mathrm{NATS}(+)=-0.8\%\;[-1.4,\,-0.3] $ and $ \mathrm{NATS}(-)=-0.8\%\;[-1.4,\,-0.2] $, indicating that, conditional on a change, both higher-score and lower-score shifts are more frequent when flipping male$\to$female than female$\to$male, consistent with the positive DTS in both directions when only the sex field is flipped and no textual information is used for triage.

Net asymmetric indices are consistent with the aforementioned skews. In the Bordeaux University Hospital ED full setting, $ \mathrm{NATS}(+)=-1.1\%\;[-1.4,\,-0.8] $ and $ \mathrm{NATS}(-)=1.0\%\;[0.7,\,1.3] $, indicating a greater propensity for upward (less severe) scores when transforming males$\to$females than females$\to$males, and a greater propensity for downward (more severe) scores when transforming females$\to$males than males$\to$females, respectively. 

\vspace{0.2cm}

\noindent\textbf{Methodological portability test.}
On the MIMIC-IV dataset, the pattern supports methodological portability: $ \mathrm{NMD}=0.022\;[0.016,\,0.027] $, $ \mathrm{DTS}_{M\mid F}=-2.46\%\;[-3.32,\,-1.61] $ and $ \mathrm{DTS}_{F\mid M}=+1.96\%\;[1.19,\,2.67] $. Pooling the two cohorts as above yields a per-pair rate of $2.18\%\;[1.67,\,2.74]$ (bootstrap 95\% CI): for otherwise identical presentations, the female variant is about $2.2$ percentage points more likely than the male variant to receive a higher (less severe) predicted triage score, a directionally consistent but larger effect than in the Bordeaux cohort. The PDR was $28.16\%\;[27.7,\,28.6]$, indicating a higher overall rate of score changes when transformations are applied. Net asymmetries were $ \mathrm{NATS}(+)=-1.49\%\;[-2.24,\,-0.76] $ and $ \mathrm{NATS}(-)=+2.93\%\;[2.21,\,3.66] $, again indicating more upward (less severe) changes for male$\to$female transformations than female$\to$male and more downward (more severe) changes for female$\to$male than male$\to$female.

\begin{figure}[H]
\centering
\caption*{\textbf{Gender-bias metrics across datasets.}}
\includegraphics[width=0.99\linewidth]{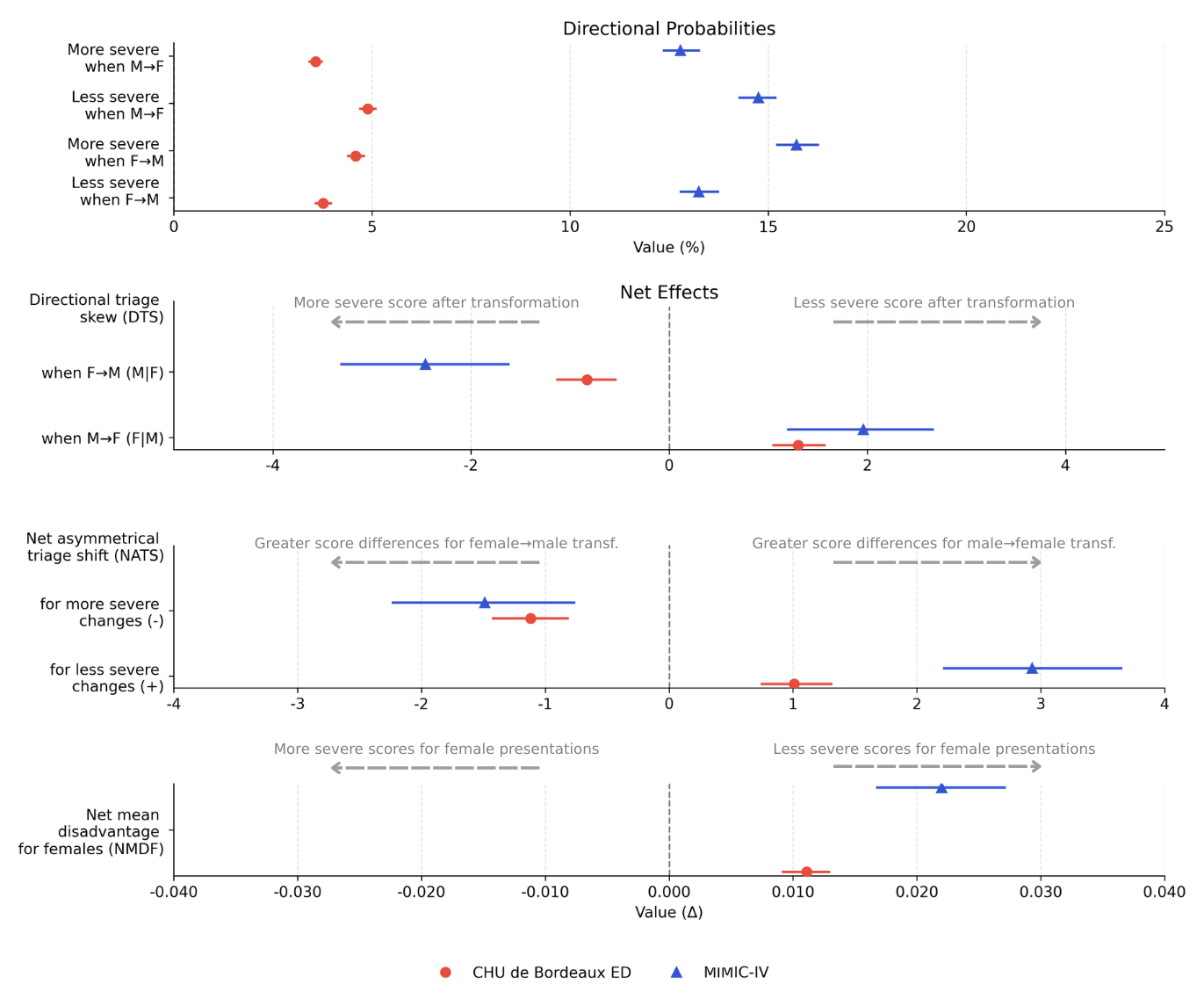}
\caption{Forest plots summarize point estimates and 95\% bootstrap confidence intervals for the Bordeaux Universtiy Hospital ED and MIMIC-IV test sets.}
\label{fig:bias_comparison}
\end{figure}

The combination $ \mathrm{DTS}_{F\mid M}>0 $ and $ \mathrm{DTS}_{M\mid F}<0 $ in both datasets quantifies a consistent asymmetry whereby female presentations are more likely than male presentations to receive higher predicted triage scores for the same clinical content. Figure~\ref{fig:bias_comparison} visualizes these directional imbalances, and Table~\ref{tab:bias_metrics_side_by_side} reports all estimates with 95\% bootstrap confidence intervals.

Stratifying the paired-comparison agreement by triage level reveals a clear acuity dependence in both datasets. Effects are negligible at the most critical level and data-sparse at the trivial end (MIMIC Triage level 4). The largest divergence occurs at intermediate acuity (Triage level 3), where transforming female→male is more often judged ``more severe" than male→female (MIMIC: \( \mathrm{OR}=0.81,\ 95\%\ \mathrm{CI}\ [0.75,\,0.88] \); Bordeaux Univ. Hosp. ED: \( \mathrm{OR}=0.62,\ 95\%\ \mathrm{CI}\ [0.55,\,0.69] \)). Full percentage breakdowns and odds ratios by level are reported in Supplementary Tables 3--6.

\begin{landscape}
\vfill
\begin{table}
\centering
\begin{tabularx}{\linewidth}{@{}l>{\centering\arraybackslash}X>{\centering\arraybackslash}X>{\centering\arraybackslash}X|>{\centering\arraybackslash}X@{}}
\toprule
 & \multicolumn{3}{c}{\textbf{Bordeaux University Hospital ED}} & \textbf{MIMIC-IV} \\
\cmidrule(lr){2-4}\cmidrule(lr){5-5}
\textbf{Metric} & \textbf{Full} & \textbf{Text-Iso} & \textbf{Tab-Iso} & \textbf{Full} \\
\midrule
Pairwise Disagreement Rate (PDR) & $8.4\%$ $[8.2\%$--$8.6\%]$ & $18.7\%$ $[18.4\%$--$19.0\%]$ & $31.9\%$ $[31.6\%$--$32.3\%]$ & $28.16\%$ $[27.70\%$--$28.61\%]$ \\
\midrule
\multicolumn{5}{@{}l}{\textit{Directional Probabilities}} \\
More severe when $M \to F$ ($\mathbb{P}^{\downarrow}_{\text{M}\to\text{F}}$) & $3.6\%$ $[3.4\%$--$3.8\%]$ & $14.4\%$ $[14.0\%$--$14.7\%]$ & $14.5\%$ $[14.1\%$--$14.8\%]$ & $12.78\%$ $[12.34\%$--$13.28\%]$ \\
Less severe when $M \to F$ ($\mathbb{P}^{\uparrow}_{\text{M}\to\text{F}}$)  & $4.9\%$ $[4.7\%$--$5.1\%]$ & $4.1\%$ $[3.9\%$--$4.3\%]$ & $18.2\%$ $[17.8\%$--$18.6\%]$ & $14.75\%$ $[14.25\%$--$15.21\%]$ \\
More severe when $F \to M$ ($\mathbb{P}^{\downarrow}_{\text{F}\to\text{M}}$) & $4.6\%$ $[4.4\%$--$4.8\%]$ & $15.6\%$ $[15.2\%$--$16.1\%]$ & $13.6\%$ $[13.3\%$--$14.0\%]$ & $15.71\%$ $[15.20\%$--$16.28\%]$ \\
Less severe when $F \to M$ ($\mathbb{P}^{\uparrow}_{\text{F}\to\text{M}}$)  & $3.8\%$ $[3.6\%$--$4.0\%]$ & $3.3\%$ $[3.1\%$--$3.5\%]$ & $17.4\%$ $[17.0\%$--$17.8\%]$ & $13.25\%$ $[12.77\%$--$13.76\%]$ \\
\midrule
\multicolumn{5}{@{}l}{\textit{Net Effects}} \\
Directional Triage Skew for M$\to$F ($\mathrm{DTS}_{F\mid M}$)  & $1.3\%$ $[1.0\%$--$1.6\%]$ & $-10.3\%$ $[-10.7\%$--$-9.9\%]$ & $3.8\%$ $[3.2\%$--$4.3\%]$ & $1.96\%$ $[1.19\%$--$2.67\%]$ \\
Directional Triage Skew for F$\to$M ($\mathrm{DTS}_{M\mid F}$)  & $-0.8\%$ $[-1.1\%$--$-0.5\%]$ & $-12.3\%$ $[-12.8\%$--$-11.9\%]$ & $3.8\%$ $[3.2\%$--$4.4\%]$ & $-2.46\%$ $[-3.32\%$--$-1.61\%]$ \\
Downward Net Asymmetric Triage Shift (NATS($-$)) & $1.0\%$ $[0.7\%$--$1.3\%]$ & $1.2\%$ $[0.7\%$--$1.8\%]$ & $-0.8\%$ $[-1.4\%$--$-0.3\%]$ & $2.93\%$ $[2.21\%$--$3.66\%]$ \\
Upward Net Asymmetric Triage Shift (NATS($+$)) & $-1.1\%$ $[-1.4\%$--$-0.8\%]$ & $-0.8\%$ $[-1.0\%$--$-0.5\%]$ & $-0.8\%$ $[-1.4\%$--$-0.2\%]$ & $-1.49\%$ $[-2.24\%$--$-0.76\%]$ \\
Net Mean Difference (NMD) & $0.011$ $[0.009$--$0.013]$ & $-0.0024$ $[-0.0056$--$0.0006]$ & $0.0055$ $[0.0011$--$0.01]$ & $0.022$ $[0.0167$--$0.0272]$ \\
\bottomrule
\end{tabularx}
\caption{Side-by-side comparison of gender bias metrics between the Bordeaux University Hospital ED and the MIMIC-IV datasets. Bordeaux results are reported for three transformation conditions (Full, Text-isolated, Tabular-isolated); MIMIC-IV reports the Full transformation only. Brackets show 95\% confidence intervals from 1,000 bootstrap iterations.}
\label{tab:bias_metrics_side_by_side}
\end{table}
\end{landscape}

\begin{figure}[H]
\centering
\caption*{\textbf{Effect of pre-training on measured gender bias.}}
\includegraphics[width=0.99\linewidth]{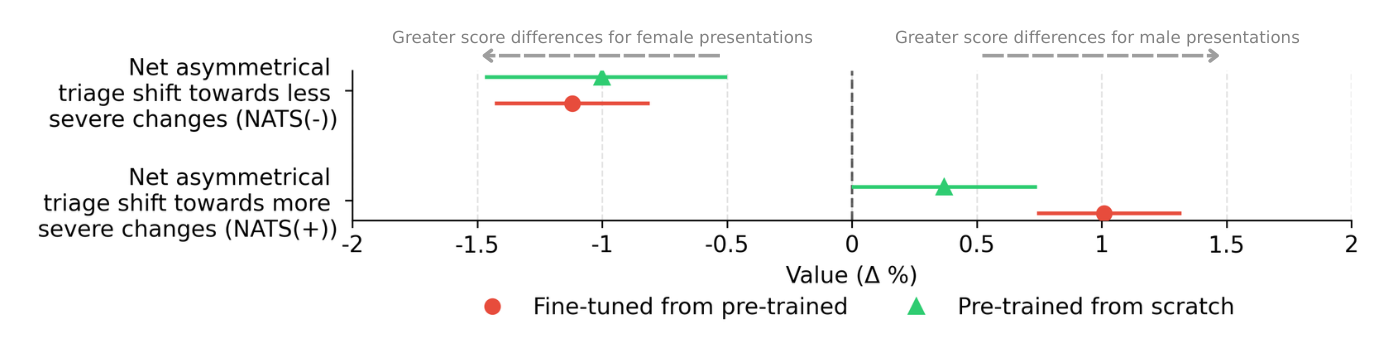}
\caption{The figure compares NATS(-) and NATS(+) for a model fine-tuned from a general-domain checkpoint versus a model trained from scratch on ED data.}
\label{fig:bias_comparison_pretrain}
\end{figure}

\noindent\textbf{Are pre-trained models more biased?}
To directly evaluate the impact of pre-training on gender bias in emergency triage, we compared two model variants on the Bordeaux University Hospital ED test set: an LLM that has been pre-trained on a diverse corpus (Mistral NeMo 12B) and subsequently fine-tuned for triage prediction (``fine-tuned from pre-trained"), and a BERT-based architecture which was trained only on data from the train partition of the Bordeaux University Hospital ED dataset (``pre-trained from scratch"). Figure~\ref{fig:bias_comparison_pretrain} visualizes this comparison across two metrics: NATS(-) and NATS(+).

For differences in less severe changes $(\text{NATS(-)})$, both models are essentially identical: $\text{NATS(-)}_{\text{Fine-tuned}}=-1.12\%\;[-1.43,\,-0.81]$ versus $\text{NATS(-)}_{\text{Pre-trained}}=-1.00\%\;[-1.47,\,-0.50]$, with overlapping intervals and a negligible difference ($\approx 0.12$ percentage points). For differences in more severe changes  $(\text{NATS(+)})$, the fine-tuned model is somewhat larger: $\text{NATS(+)}_{\text{Fine-tuned}}=+1.01\%\;[+0.74,\,+1.32]$ versus $\text{NATS(+)}_{\text{Pre-trained}}=+0.37\%\;[+0.00,\,+0.74]$; this suggests a modest increase but is borderline given that the pre-trained model's lower bound touches zero. 

This small sensitivity evaluation suggest that while general-domain pre-training may slightly amplify net gender bias in some scenarios, the primary driver of gendered triage disparities is not the pre-training procedure itself, but the underlying structure of gender cues in clinical input data and the language patterns learned during fine-tuning.

\vspace{0.2cm}
\noindent\textbf{Sex-scrubbed baseline model.} 
To test whether the observed gender asymmetries are mediated by explicit sex/gender markers rather than correlated proxies, we retrained Mistral NeMo 12B on the MIMIC-IV training partition after removing all gender information: gendered language in clinical notes was replaced with neutral equivalents using the same LLM-based rewriting pipeline, and the tabular sex field was removed. Table~\ref{tab:neutralized} compares the original and neutralized models on the MIMIC-IV test set ($n = 38{,}216$). The prompt for this transformation can be found in Supplementary Note~3.
 
\begin{table}[h]
\centering
\begin{tabular}{lccccc}
\toprule
 & M--F gap & 95\% CI & $p$-value & Within-1 acc. & $\kappa_w$ \\
\midrule
Fine-tuned with gender & $-0.0332$ & $[-0.0453,\ -0.0205]$ & $6.4 \times 10^{-7}$ & 99.2\% & 0.603 \\
Fine-tuned neutralized & $-0.0005$ & $[-0.0128,\ +0.0128]$ & $0.93$ & 99.5\% & 0.627 \\
\bottomrule
\end{tabular}
\caption{Sex-scrubbed baseline comparison on MIMIC-IV. M--F gap is the difference in mean predicted triage score between male and female patients; the 95\% confidence interval is a bootstrap interval on this mean difference (1{,}000 iterations). The Mann--Whitney $U$ test evaluates whether male and female prediction distributions differ significantly.}
\label{tab:neutralized}
\end{table}
 
The central finding is that removing gender information eliminated the between-sex prediction gap: the M--F difference collapsed from $-0.0332$ (95\% CI $[-0.045,\ -0.021]$; Mann--Whitney $p = 6.4 \times 10^{-7}$) to $-0.0005$ (95\% CI $[-0.013,\ +0.013]$; $p = 0.93$). The bootstrap interval for the original model excludes zero, while the interval for the neutralized model spans it, providing converging evidence that the observed disparity is mediated by explicit gender markers. Performance metrics remained comparable ($\kappa_w$: $0.603 \to 0.627$), confirming that the neutralized model maintains adequate predictive quality. We note that this neutralization analysis was conducted on MIMIC-IV only. It therefore supports the interpretation that explicit sex/gender markers contribute to the observed model asymmetries in that setting, but it should not be assumed to transfer to the Bordeaux cohort, where an analogous retraining-on-neutralized-inputs experiment was not performed; extending this baseline to the French data is a natural next step.
\vspace{0.2cm}

\begin{figure}[H]
\centering
\includegraphics[width=0.99\linewidth]{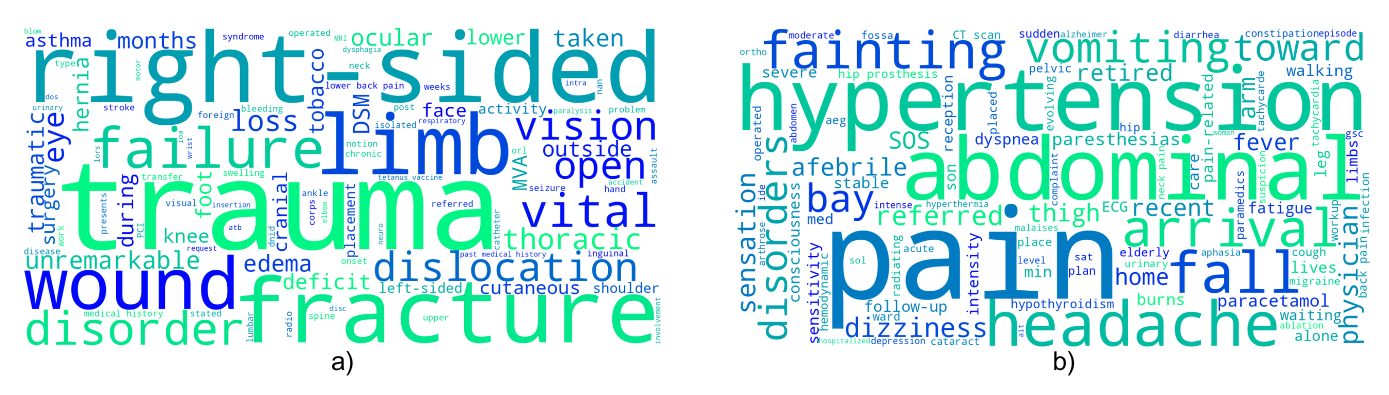}
\caption{\textbf{Word clouds of clinical presentations associated with differential triage in the Bordeaux University Hospital ED dataset (Translated to English)}. (a) M$\to$F transformations receiving less critical triage after transformation: prominent terms include trauma-related presentations. (b) F$\to$M transformations receiving more critical triage after transformation: prominent terms include pain and non-specific presentations.}
\label{fig:wordclouds_chu}
\end{figure}

\noindent\textbf{Characterizing differentially triaged presentations.}
To understand what types of clinical presentations are most affected by gender-based differential triage, we examined the clinical notes of cases where predictions changed after gender transformation. Specifically, we identified cases where: (1) originally male patients received less severe triage scores when presented as female (M→F less severe), and (2) originally female patients received more severe triage scores when presented as male (F→M more severe). Figure~\ref{fig:wordclouds_chu} presents word clouds of the most frequent terms in the clinical notes for each category in the Bordeaux University Hospital ED dataset.

The patterns revealed by these word clouds align with documented clinical disparities. Presentations involving pain (abdominal, chest), gastrointestinal symptoms (nausea, vomiting), and non-specific complaints (weakness, fever) show the strongest gender-related differential treatment. These are precisely the presentations where prior research has documented systematic undertriage of women, particularly for conditions like acute coronary syndrome where women's ``atypical'' presentations are often assigned lower acuity than equivalent male presentations \cite{mehilli2020coronary}. These word clouds are exploratory: they reflect the frequency of individual terms rather than structured clinical categories, and we therefore treat them as hypothesis-generating rather than as a basis for clinical conclusions. A more clinically interpretable characterization by complaint group or symptom cluster is a target for future work (see Discussion). A corresponding analysis for the MIMIC-IV dataset is provided in Supplementary Figure 3.

\begin{figure}[H]
\centering
\caption*{\textbf{DTS by patient--nurse sex (95\% confidence intervals).}}
\includegraphics[width=0.85\linewidth]{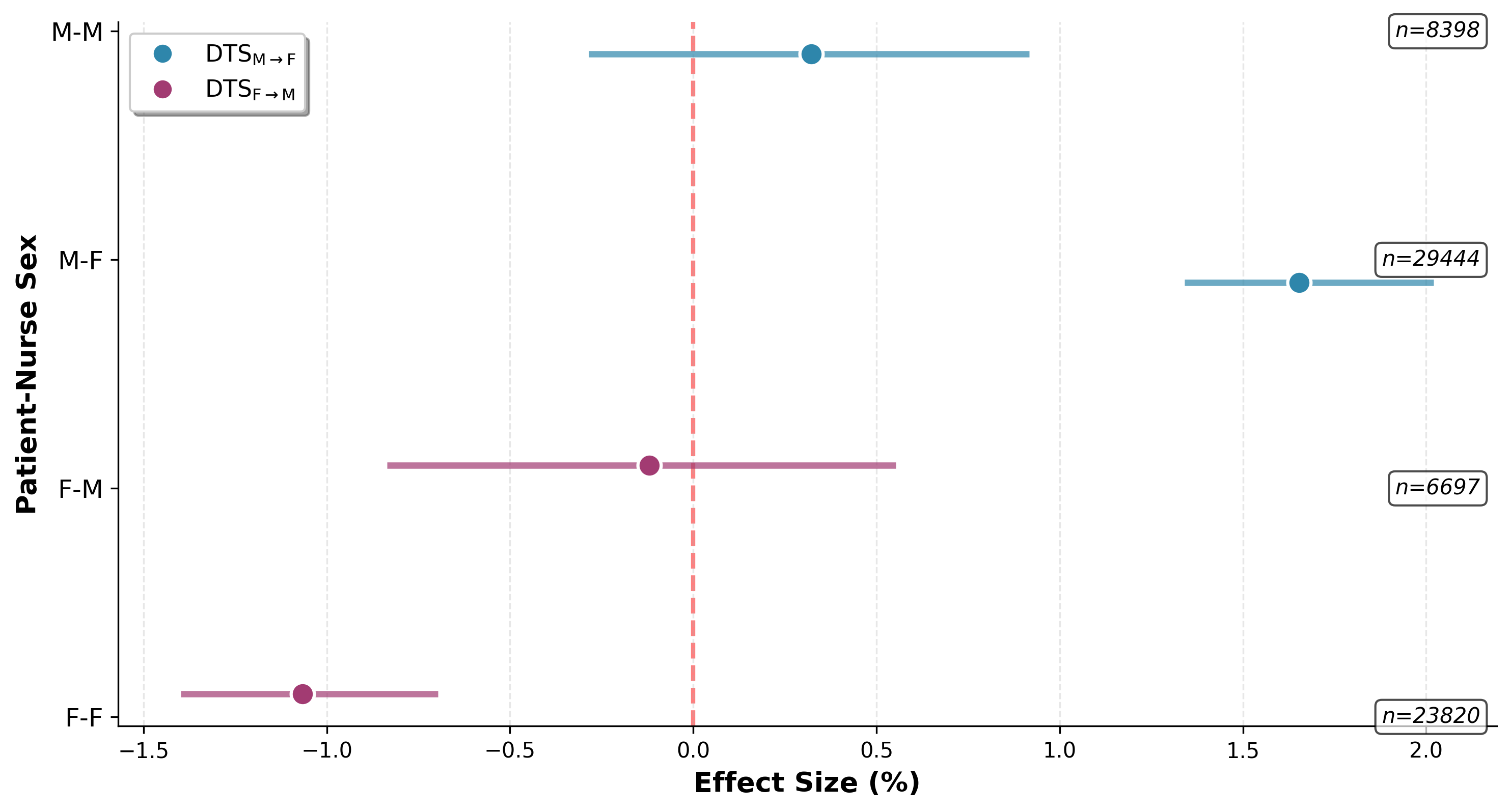}
\caption{DTS is shown for each patient--nurse sex combination in the Bordeaux University Hospital ED dataset (M-M: n=8,398; M-F: n=29,444; F-M: n=6,697; F-F: n=23,820), with 95\% confidence intervals.}
\label{fig:nurse_patient_sex}
\end{figure}

\vspace{0.2cm}
\noindent\textbf{Role of triage nurse sex.}
The Bordeaux University Hospital ED dataset includes information about the sex of the triage nurse who assessed each patient, enabling us to examine whether bias patterns vary by nurse-patient sex concordance. Figure~\ref{fig:nurse_patient_sex} presents DTS metrics stratified by the four possible patient-nurse sex combinations.

The results reveal a striking interaction between patient and nurse sex. Female nurses show the strongest bias effects: when triaging male patients (M-F), $\text{DTS}_{M\to F}$ is strongly positive (+1.6\%), indicating that male presentations tend to receive less severe scores when appearing female. Conversely, when female nurses triage female patients (F-F), $\text{DTS}_{F\to M}$ is strongly negative (-1.0\%), indicating that female presentations tend to receive more severe scores when appearing male, in other words, female patients receive less critical scores when triage is performed by female nurses relative to how they would be triaged if presenting as male.

Male nurses (M-M and F-M conditions) show minimal differential treatment, with both DTS metrics close to zero regardless of patient sex. This pattern suggests that gender-based differential triage may be more pronounced in same-sex nurse-patient dyads or may reflect different decision-making patterns between male and female nurses. We caution that this analysis does not establish any psychological mechanism: interpretations such as in-group/out-group dynamics or implicit-bias pathways are offered as hypotheses for future work rather than as conclusions supported by the present design. These findings warrant further investigation to understand the mechanisms underlying nurse-patient sex interactions in triage decisions.

\section*{Discussion}\label{discussion}

This study extends prior work by applying LLMs in a large-scale paired-comparison analysis to probe documented sex/gender-related asymmetries in emergency triage decisions. Its primary contribution is methodological: it demonstrates the feasibility of using a fine-tuned LLM as a scalable, documentation-level probe of recorded decisions. Applied at scale, the probe reveals consistent directional asymmetries in model-predicted triage scores under gender-swapped perturbations, detected across two large emergency department datasets from different countries and healthcare systems. These asymmetries are small in magnitude; they are consistent with the hypothesis that sex/gender cues influence triage assignment, but their clinical relevance is not established, and clinician-anchored validation is needed before attributing them to bedside decision-making.

A notable finding is a consistent directional skew in how identical presentations are triaged after gender flips, summarized by the Directional Triage Skew (DTS). In the full Bordeaux University Hospital ED setting, $\text{DTS}_{M\mid F}=+1.3\%\,[1.0,\,1.6]$ and $\text{DTS}_{F\mid M}=-0.8\%\;[-1.1,\,-0.5]$. Pooled across cohorts into a single per-pair rate, model-predicted scores were about $1.1\%\;[0.89,\,1.29]$ more likely to be less severe when the same documented clinical content was presented as female rather than male.

At the level our design can resolve, this asymmetry indicates a small but consistent documentation-level sensitivity of model predictions to sex/gender markers, oriented in a direction that would disadvantage female presentations. Whether it reflects a clinically relevant disadvantage remains hypothetical: the magnitude is small, and in the absence of clinician-anchored or outcome-based validation we cannot establish that a gender-related heuristic operates at the bedside or that it alters care. We therefore state the following as a conditional implication rather than as a claim about current practice. If these documentation-level patterns were to reflect clinical practice, the implications would be significant: in emergency medicine, where minutes can determine outcomes, systematic undertriage could contribute to delayed diagnosis and treatment, which has in turn been linked to sex differences in outcomes for conditions like acute coronary syndrome and stroke \cite{mehilli2020coronary}.

To give a rough sense of scale, the pooled documentation-level rate can be projected onto national emergency volumes, though we do so with reservation and treat the result as an illustration only. France records roughly 20.9 million annual emergency visits \cite{drees2024urgences}, of which women account for just under half (we use 48\%, as French emergency attendance is consistently slightly male-majority); applied mechanically, a $1.1\%$ differential is on the order of 110{,}000 additional lower-severity assignments for women per year (illustrative range roughly 90{,}000 to 130{,}000 from the confidence interval of the pooled rate). We stress that this figure multiplies a very small marginal model effect by a very large denominator and is therefore easily over-interpreted: it is not a count of harmed patients, and it assumes, without evidence, that a documentation-level, model-predicted shift would transfer unchanged to bedside care.

Critically, this projection is not anchored to any tangible care difference. Clinically interpretable disparity studies calibrate magnitude against concrete endpoints such as treatment received, waiting time, downstream outcomes, or expert-defined mistriage \cite{onal2022throughput, lopez2021poisoning, chen2008gender, amy2019sex}; our documentation-level, model-predicted signal is smaller in kind and is not directly comparable to those care-anchored measures. Translating the detected shifts into care-relevant units, for instance by mapping one-level downgrades onto the system-defined target times to medical contact in the CIMU/FRENCH scale, or by linking affected cases to measured waiting times, resource use, and outcomes, is precisely what the validation study outlined below would require. Until such anchoring is performed, the national figure should carry little practical weight, and we retain it only as a coarse indication of the population scale at which even small documentation-level effects would operate.

The mean-difference metric is directionally consistent with the skew. Female presentations received an average predicted triage score $0.011\,[0.009\text{--}0.013]$ points higher (less urgent) than identical male presentations. This magnitude is very small on the five-level scale and reaches statistical significance largely because of the very large sample. Whether it would translate into meaningful differences in wait times or resource allocation depends on the triage system and remains to be established empirically. Together, these findings underscore ethical considerations for responsible AI use in clinical pathways and the need for governance around fairness auditing \cite{char2018toward}.

Our paired-comparison framework contributes to the field of bias detection in several ways. First, by training an LLM (Mistral NeMo 12B) to approximate human triage decisions as recorded in clinical documentation, we obtain a model that reflects documented current clinical practice with agreement comparable to reported human inter-rater levels. This approach sidesteps the challenge of defining ``correct'' triage assignments, instead focusing on detecting inconsistencies in how patient gender influences predictions conditioned on documented inputs. Second, our modality-specific analysis reveals that bias operates through multiple channels. The high sensitivity when transforming only tabular gender indicators suggests that explicit gender markers strongly influence triage, while the different patterns observed with text-isolated transformations indicate that implicit gender cues in clinical narratives also contribute, albeit through different mechanisms. Third, the use of asymmetric bias metrics (NATS) provides insights beyond simple parity measures. Traditional fairness metrics might miss these directional effects, which reveal not just that predicted outcomes differ by gender, but that the model responds asymmetrically to gender transformations, an asymmetry that, given the model was trained on documented decisions, may reflect patterns in the underlying recorded human decision-making.

A critical question for any LLM-based bias detection framework is whether observed disparities reflect genuine patterns in the underlying data or are artifacts of model behavior. Our stratified analysis by triage nurse sex provides supportive evidence for the former interpretation. If the observed gender asymmetries were purely model artifacts, arising from noise, hallucinations, or arbitrary learned associations, we would not expect them to vary systematically with characteristics of the decision-maker (the triage nurse). Yet we observe precisely such systematic variation: female nurses exhibit the strongest effects, with pronounced asymmetries in both the M-F and F-F nurse-patient dyads, while male nurses show minimal differential treatment regardless of patient sex (Figure~\ref{fig:nurse_patient_sex}). Because triage nurse sex was part of the full training context provided to the model (as described in the Methods), alongside the other tabular variables from the clinical record, the systematic variation across nurse-patient sex combinations indicates that the model faithfully reproduced the differential decision patterns associated with each nurse-patient dyad as recorded in the data. We frame possible explanations, such as in-group/out-group dynamics or implicit-bias pathways, as hypotheses motivated by prior literature rather than as mechanisms established by the present design; the analysis cannot adjudicate among them. Taken cautiously, this pattern offers methodological reassurance that the framework detects signals present in documented clinical data rather than manufacturing spurious effects, while the psychological interpretation of those signals remains a question for future work.

While this case study focuses on emergency triage, the proposed paired-transformation framework is domain-agnostic and can be applied wherever bias testing and auditing is conducted. Given access to decision inputs (structured and/or textual) and a well-specified transformation for the protected attribute(s), the same methodology can audit human resources decisions (e.g., screening, hiring, promotion), academic decisions (e.g., admissions, grading, fellowship selection), and justice-related decisions (e.g., risk assessment, pretrial release, sentencing). The approach does not require defining an absolute ground truth; instead, it detects directional asymmetries in how outcomes change under protected-attribute transformations, making it directly compatible with existing discrimination testing practices. Moreover, our cross-country and cross-language portability test suggests the framework can be applied across institutional and linguistic settings, though the structural differences between the Bordeaux and MIMIC datasets (discharge notes vs. triage notes, CIMU vs. ESI scales, different severity constructs at intermediate levels) mean that the MIMIC analysis tests methodological transferability rather than clinical generalizability of the same disparity.

A strength of this within-pair design bears directly on a natural objection: that men presenting to the emergency department may, on average, be more acutely ill than women, so that sex might carry legitimate prognostic information rather than reflect bias. Because each comparison holds the documented clinical content fixed and flips only the sex/gender markers, population-level differences in baseline morbidity between men and women cannot drive the within-pair asymmetry we measure. The contrast is between two presentations of the same case, not between two different patients, so any such difference is controlled by construction. This does not rule out that the documented content itself is shaped by sex/gender, but it does separate the asymmetry we report from population-level differences in who presents more severely ill.

Interpreting the direction and clinical meaning of category shifts requires care, because the two triage systems encode acuity differently. The ESI used in MIMIC-IV incorporates a substantial expected-resource-use component, particularly at intermediate and lower acuity levels, so a shift from ESI~3 to ESI~4 may primarily reflect anticipated resource needs rather than urgency per se. The French CIMU/FRENCH scale is closer to an urgency- or time-to-contact-oriented system (in the spirit of the Manchester Triage System), in which a downward category shift more directly implies a longer target time to medical contact and thus a potential delay in treatment. The same nominal one-level shift can therefore carry different clinical implications across the two systems, which is a further reason to treat the MIMIC-IV analysis as a portability test rather than a clinical replication. For the same reason, not all shifts are equally consequential: a shift between adjacent low-acuity categories for a stable, self-limiting complaint differs substantially from a shift affecting potentially unstable presentations such as chest pain, abdominal pain, dyspnoea, neurological symptoms, or sepsis-like states, where even a single-level downgrade could plausibly affect time-critical care. Our aggregate metrics do not by themselves separate these cases, and the acuity-stratified results and the exploratory presentation-level word clouds should be read with this distinction in mind.

Relatedly, the directionality of a shift does not map cleanly onto harm. Overtriage is not necessarily an error but is, in many triage systems, a deliberate safety margin: assigning a higher acuity than strictly required reduces the risk of missing a deteriorating patient. Consequently, a shift toward higher severity for one sex is not, in clinical terms, the simple mirror image of a shift toward lower severity for the other, and movements toward higher- or lower-severity categories need not carry symmetric safety implications. We therefore read the observed asymmetries primarily as evidence that documented sex/gender cues move predictions, and only secondarily, and tentatively, as evidence about the direction of potential clinical risk, which depends on where on the acuity range a shift occurs and on the presentation involved.

It is important to distinguish our paired-perturbation approach from formal causal inference methods. While our framework draws conceptual inspiration from counterfactual reasoning (asking ``what would the triage score have been if this patient had presented as the opposite gender?''), we do not claim to estimate causal effects in the technical sense used in the causal inference literature \cite{kusner2017counterfactual}. Instead, our approach is best understood as a \textit{sensitivity analysis} or \textit{perturbation study}: we systematically modify gender-related attributes in patient records and measure how model predictions respond to these perturbations. The resulting metrics quantify the model's sensitivity to gender information, which (given that the model was trained to emulate human decisions) provides evidence about patterns in the underlying human decision-making process. We interpret consistent directional asymmetries as \textit{associations consistent with bias} rather than as proof of causal relationships. This framing aligns with audit and fairness testing methodologies that examine outcome disparities without making strong causal claims. We also acknowledge the literature on causal estimands for ordinal outcomes, including sharp bounds and distributional approaches \cite{lu2018ordinal}, which provide principled frameworks for effect identification when outcomes are ordered categories. Our objective, however, is not causal identification but rather sensitivity testing: quantifying how predictions respond to gender perturbations as descriptive evidence of differential treatment patterns.

Several limitations warrant consideration. First, there is an inherent ``multimodality gap'' between the information available to our models and the full set of cues used in real triage encounters. Our analysis relies on recorded clinical data, whereas human triage integrates additional modalities such as speech prosody, tone, visible distress, agitation, and overall demeanor. These non-verbal signals can influence perceived urgency but are not captured in our inputs, so our estimates reflect asymmetries present in documented information rather than the complete sensory context of triage. Second, while our LLM approximates documented human triage decisions, it necessarily inherits any asymmetries present in the training data. Our analysis detects relative differences in how gender influences triage but cannot determine absolute ``correct'' triage levels. Third, our binary gender framework, while enabling systematic analysis, does not capture the full spectrum of gender identity and its intersection with other demographic factors. Additionally, our approach examines associations between gender and triage outcomes rather than establishing strict causal relationships, as the gender transformation process cannot account for all potential confounders or ensure that the modified records perfectly represent how actual patients of the transformed gender would present. Fourth, while our observed effects are consistent and directionally stable across analyses, there remains the possibility that the LLM itself may \textit{amplify or distort} certain biases present in the training data differently than humans; the complex, non-linear nature of transformer architectures could magnify subtle patterns in ways that do not directly correspond to the original human decision-making processes. We are currently investigating this phenomenon, including through latent space exploration, to better understand \textit{how} the model reproduces these patterns and to quantify any potential amplification or distortion effects. Fifth, although decoding at inference time is deterministic for a fixed trained model (temperature $= 0$, greedy decoding), we did not conduct multi-seed fine-tuning robustness checks to verify that the observed effects are stable across random initializations and training runs; the convergence of the classic counterfactual and gender-neutralization frameworks on consistent directional conclusions partially mitigates this concern but does not fully replace a formal stability analysis. Sixth, blinded clinician re-triage of sex-neutralized or gender-swapped notes would represent a reasonable gold-standard comparator for validating the clinical significance of the detected documentation-level patterns; our findings strongly motivate such a study as a critical next step.

A responsible next validation step follows directly from these limitations. Establishing whether the documentation-level asymmetries we detect correspond to clinically meaningful triage differences would ideally involve a study in which clinicians, blinded to sex/gender, re-triage matched original and sex/gender-swapped notes, so that prediction shifts can be compared against expert human judgment rather than against documented labels alone. Such a study should, where feasible, link the affected cases to downstream outcomes or measured resource use; stratify by clinically relevant presentation groups (for example chest pain, abdominal pain, dyspnoea, neurological, and sepsis-like presentations) rather than relying on aggregate metrics or word-frequency summaries; and evaluate explicitly whether the triage shifts that occur would plausibly alter waiting times, diagnostic pathways, or treatment timing. A design of this kind, ideally paired with an analogous gender-neutralization retraining experiment on the French cohort, would move the evidence from documentation-level sensitivity toward the bedside-level inference that our current data cannot support.

Future work should extend this framework to examine intersectional biases, particularly the interaction of gender with race, age, and socioeconomic status. Additionally, prospective studies could evaluate whether awareness of these biases, facilitated by our detection framework, leads to more equitable triage practices.

Our findings should not be interpreted as evidence of conscious bias or as direct proof of bedside clinician behavior. Rather, they may indicate that documented triage records contain patterns consistent with systematic sex/gender-based asymmetries. These documentation-level patterns may reflect the influence of implicit biases and systemic factors, but clinician-anchored validation, such as blinded re-triage of sex-neutralized notes or outcome-based calibration, is needed to determine the extent to which they correspond to actual bedside decision-making. Women's symptoms are often described as ``atypical'', while their vital signs and other clinical measures are frequently assessed against reference standards that are not sufficiently adapted to them. Recent studies have begun to directly compare the performance of AI tools against human triage nurses \cite{santoro2024human, aydin2025evaluating, kocak2024transforming, nover2025comparing, lindner2025performance}. Whether these studies conclude that AI tools outperform human experts or vice versa, the primary evaluation criterion remains performance (e.g., accuracy, concordance). However, focusing solely on performance overlooks the critical risk of bias amplification inherent in these systems. Our work complements this performance-centric view by providing a necessary supplementary criterion: evaluating whether the underlying data and model behavior are sufficiently ``reliable'' and unbiased to ensure that these tools, regardless of their raw performance, can be deployed safely and ethically.

This work demonstrates how AI can serve as a tool for healthcare equity, not by replacing human judgment, but by helping us understand and address its limitations. By training an LLM to approximate human triage decisions as recorded in documentation and then systematically evaluating its behavior on gender-swapped patient presentations, we have identified consistent documentation-level patterns of sex/gender-associated asymmetry, observed in a large French cohort and reproduced in MIMIC-IV as a cross-language, cross-system methodological portability test. With appropriate validation, a framework such as this could in future support preliminary, documentation-level screening that helps identify specific clinical presentations or contexts where sex/gender-associated asymmetries are most pronounced, and so generate hypotheses for closer study. We caution that it is not yet an operational fairness audit of triage practice: using it to draw conclusions about bedside care would first require blinded clinician re-triage, chart review, or outcome-based calibration. Should such validation support the documentation-level signals, the modality-specific analysis could inform targeted interventions, for instance, modifying electronic health record interfaces to minimize the salience of demographic information during triage assessment, or developing training programs that address implicit biases in interpreting clinical narratives. A framework such as this one is best understood as a diagnostic tool that can point toward candidate recommendations for debiasing, which must ultimately be developed through expert clinical consensus. To this end, the results of this study are currently being shared and discussed with clinicians at the Bordeaux University Hospital to help inform training protocols for triage nurses. The broader ambition is to leverage these findings, after appropriate validation, to collaborate with national bodies, such as the French National College of Emergency Physicians (SFMU), toward national recommendations for more equitable emergency care. Taken together, we present this work as a methodological feasibility study: a credible contribution to documentation-level fairness auditing and a source of hypotheses for real-world triage bias. Consistent with this framing, the small size of the effects we detect is itself part of the finding: the probe is sensitive enough to surface a consistent, reproducible sex/gender-marker signal, while the signal it surfaces is small enough that its clinical significance cannot be asserted without the validation we outline above. By making visible patterns in documented decisions that may be influenced by sex/gender, we take a step toward emergency departments where a patient's urgency for care is determined solely by their clinical need.

\section*{Methods}\label{methods}

Our study utilizes two distinct, large-scale ED datasets to train our models and evaluate gender bias: a primary dataset from the Bordeaux University Hospital (\textit{CHU de Bordeaux}), in France, and a subset of the publicly-available MIMIC-IV database in the United States. Both institutions function as tertiary referral centers with full trauma capabilities, ensuring comparable scope of emergency care despite differences in language, patient populations, and documentation standards. However, structural differences in healthcare financing likely influence patient visit patterns, particularly the proportion of non-urgent presentations (ESI/CIMU level 5). In France's public system, a non-admitted ED visit is billed at a fixed national rate (known as the \textit{Forfait Patient Urgences}). This fee is covered by the national health insurance for all individuals affiliated with it, with any remaining balance typically reimbursed by complementary insurance, resulting in minimal or no out-of-pocket costs. In contrast, in the United States, ED visit charges vary by insurance plan, deductible status, and services provided, even in emergencies. These differences may partly explain the lower prevalence of Level 5 cases in the MIMIC-IV dataset compared to the Bordeaux University Hospital ED dataset (see Table~\ref{tab:dataset_characteristics} for a detailed distribution of triage levels and Supplementary Table 7 for distributions stratified by patient sex). Unlike the Bordeaux University Hospital ED dataset, the MIMIC-IV database includes a substantial proportion of patients assigned a triage score of 1 and, conversely, very few patients assigned a score of 5.

\begin{figure}[H]
\centering
\caption*{\textbf{Original Bordeaux University Hospital ED dataset: triage, age, and sex distributions.}}
\includegraphics[width=0.95\linewidth]{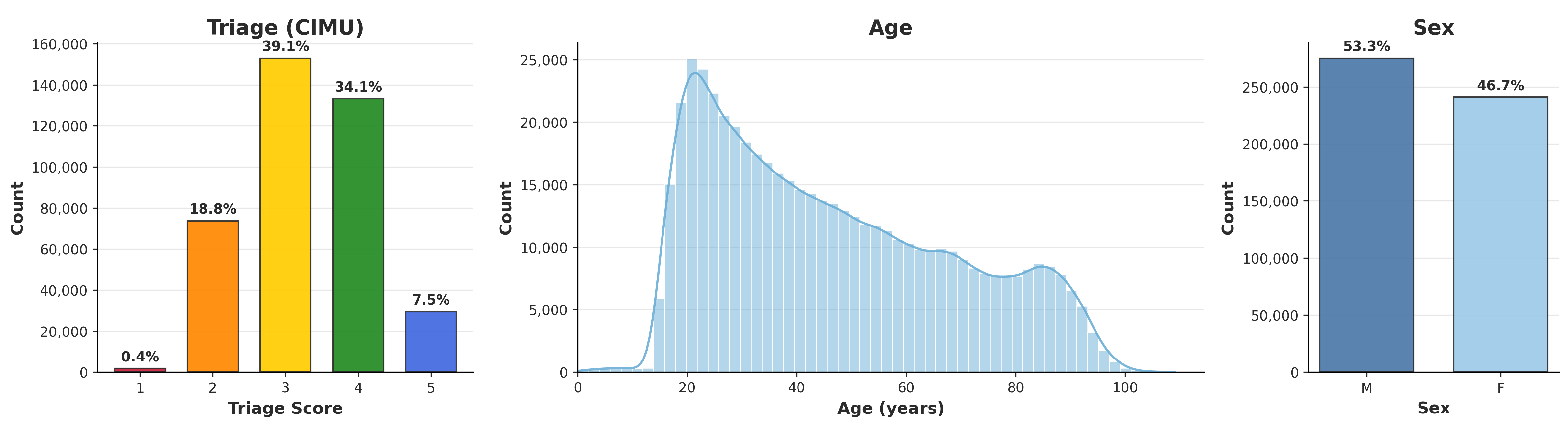}
\caption{The panels show the distributions of triage score, patient age, and patient sex/gender before any filtering steps.}
\label{fig:raw_desc}
\end{figure}

\vspace{0.2cm}
\noindent\textbf{Bordeaux University Hospital ED Dataset.}
The raw dataset used for this study comprises 520,621 admissions to the Adult Emergency Department of the Bordeaux University Hospital between January 2013 and December 2021. Each data point contains tabular variables from each admission, including an admission identifier, date and time, estimated saturation of the ED at the time of admission, patient age and gender, chief complaint, history of present illness, past medical history, vital signs (heart rate, respiratory rate, systolic and diastolic blood pressure, blood oxygen saturation, temperature, and Glasgow coma score), and the associated triage score. Additionally, information related to the nurses who performed triage was also included in the analysis, such as the gender of the triage nurse, number of years of experience at the date of triage, and whether they received specialized triage training. Figure \ref{fig:raw_desc} shows the distribution of triage scores, patient age and patient gender in the raw dataset. 

\noindent\textbf{Dataset filtering.}
An initial filtering excluded all records from before January 2016, ensuring normalization to the 5-level triage scale introduced in 2015. Samples missing patient gender, triage nurse gender, age, triage score, chief complaint, or triage notes were removed, as were patients under 18 years of age.

Gender-specific admission motives or pre-existing conditions were filtered out to ensure clinical plausibility of gender-swapped pair generation. Under our paired-comparison framework, where sex/gender is modified while holding all other clinical information constant, it would be clinically incoherent to generate pairs whose documented presentation is incompatible with the target sex (e.g., pregnancy-related conditions). We performed this filtering by examining the chief complaint, history of present illness, and past medical history. At the Bordeaux University Hospital Pellegrin ED, chief complaints are categorical and directly associated with potential ICD-10-coded diagnoses; examples of excluded categories include ``Other specified conditions associated with female genital organs and menstrual cycle'' or ``Other specified disorders of male genital organs.'' The remaining samples were screened using regular expression searches for words associated with gender-specific conditions (e.g., \textit{uter-} or \textit{prostat-}). The excluded chief-complaint classes and the complete French- and English-language keyword lists are documented in Supplementary Note~2 and Supplementary Tables~1--2, respectively.

Samples with triage score of 1 (resuscitation) were excluded because patients with immediately life-threatening conditions are often identified pre-triage and routed directly to resuscitation areas, where formal triage levels are not always documented. This exclusion avoids systematic missingness in the outcome variable and reflects real-world ED workflow.

The succession of filtering steps resulted in a sample size of 144,888 (see Supplementary Figure 1 for a detailed flowchart). This filtered dataset was randomly divided into train and test partitions of equal size (50\% each). All tabular and free-text data were then linearized into single strings containing all available information for each patient and their respective variable names.

\vspace{0.2cm}
\noindent\textbf{MIMIC-IV Dataset.}
The complete MIMIC-IV database contains data from about 540,000 hospitalizations of patients admitted to the emergency department or intensive care unit of the Beth Israel Deaconess Medical Center in Boston, Massachusetts, between 2008 and 2019. The database is composed of different modules containing different information of different units, such as the emergency department (ED) module, or the clinical notes (Note) module. 

In order to assemble a dataset which resembles the Bordeaux University Hospital ED dataset, we selected the ED, Hospitalization (Hosp) and Note modules of the MIMIC-IV database. More specifically, we selected variables from the \textit{triage} and \textit{edstays} tables from the ED module, the \textit{admissions}, \textit{patients} and \textit{providers} tables from the Hosp module and the \textit{discharge} table from the Note module. The first filtering step for this dataset was the linkage between all the aforementioned tables, only keeping samples for patients for which linkage across the three modules was possible.

We excluded samples with a triage score of five ($\sim 0.2\%$ of linked samples) in order to replicate the four-level triage task from the Bordeaux University Hospital ED dataset, and potentially avoiding a significant class-imbalance problem. As with the Bordeaux dataset, samples were screened using regular expressions containing gender-specific words which could not be reasonably transformed into gender-swapped pairs. Additionally, since the free-text notes in the MIMIC-IV database contain the complete patient discharge notes, we performed a second regular expression search for any sections in the texts which would not be relevant at triage time. 

The remaining sample, composed of about 76,432 admissions, was divided into train and test partitions of equal size (see Supplementary Figure 2 for a detailed filtering flowchart).

\begin{table}[h]
\centering
\begin{tabular}{lcc}
\toprule
\textbf{Characteristic} & \textbf{Bordeaux University Hospital ED} & \textbf{MIMIC-IV} \\
\midrule
Location & Bordeaux, France & Boston, MA, USA \\
Time Period & 2016--2021 & 2008--2019 \\
Language & French & English \\
Triage Scale & CIMU (5-level) & ESI (5-level) \\
\midrule
Total Visits & 144,888 & 76,432 \\
Age (years), mean $\pm$ SD & $46.2 \pm 22.3$ & $60.2 \pm 18.2$ \\
Male:Female Ratio & 1:0.865 & 1:1.265 \\
\midrule
\multicolumn{3}{l}{\textit{Triage Level Distribution:}} \\
\quad Level 1 (Most severe) & (Filtered) & 11.8\% \\
\quad Level 2 & 17.5\% & 48.3\% \\
\quad Level 3 & 41.2\% & 39.5\% \\
\quad Level 4 & 35.1\% & 0.4\% \\
\quad Level 5 (Least severe) & 6.2\% & (Filtered) \\
\bottomrule
\end{tabular}
\caption{Characteristics of the filtered Bordeaux University Hospital and MIMIC-IV Datasets. The Bordeaux dataset uses the CIMU triage scale and MIMIC-IV uses the ESI scale; detailed descriptions of both scales and their comparison are provided in Supplementary Note 4 and Supplementary Tables 9--10.}
\label{tab:dataset_characteristics}
\end{table}

\vspace{0.2cm}
\noindent\textbf{Model architectures.}
We compared classical machine learning and deep-learning architectures, from feature-based baselines to medium- and large-scale LLMs:
\begin{itemize}
    \item \textbf{RF/TF-IDF baseline}: Random Forest model trained on TF-IDF word vectors.
    \item \textbf{CHUBert} ($3.3 \times 10^{8}$ params): a large RoBERTa-style model pre-trained with masked language modeling on the Bordeaux University Hospital train split, then fine-tuned with a classification head for triage prediction.
    \item \textbf{MLP w/ CamemBERT embeddings} ($3.3 \times 10^{8}$ params): a multi-layer perceptron with a single hidden layer of 200 units, taking CamemBERT-derived embeddings as input.
    \item \textbf{Mistral NeMo 12B}: a 12B-parameter LLM fine-tuned for the triage task.
    \item \textbf{gpt-oss-20b}: a 20B-parameter LLM fine-tuned for the triage task.
    \item \textbf{MedGemma 27B}: a 27B-parameter medical LLM fine-tuned for the triage task.
\end{itemize}

A number of Transformed-based language models were considered for the task of automatic triage, mainly because of their capacity to encode and process large amounts of unstructured texts in multiple languages. These models were selected according on their size and performance balance based on existing benchmarks, ranging from a traditional encoder-only models to state-of-the-art large Mixture of Experts (MoE) models. 

In order to establish a baseline, a Random Forest classifier was fitted using TF-IDF vectorization on the clinical notes. The optimal hyperparameter combination for this model was found using 3-fold cross-validation on a sub-partition of the train set. Similarly, a multi-layer perceptron (MLP) with one hidden layer was trained from scratch on word embeddings produced with the CamemBERT large model.

To assess the potential impact of pre-existing biases in pre-trained LLMs, we construct a RoBERTa-style encoder entirely from scratch on the Bordeaux University Hospital ED training partition and then adapt it to triage via supervised fine-tuning. First, a byte-level BPE tokenizer is trained on all ED free text (vocabulary size 52{,}000, minimum frequency 2) with standard RoBERTa special tokens. The masked-language model is initialized randomly with a fresh \texttt{RobertaConfig} aligned to the tokenizer (\(\texttt{vocab\_size}=52{,}000\), \(\texttt{max\_position\_embeddings}=1{,}024\), \(\texttt{num\_hidden\_layers}=12\), \(\texttt{num\_attention\_heads}=12\), \(\texttt{type\_vocab\_size}=1\); dropout \(=0.1\) for both hidden and attention probabilities), yielding a \texttt{RobertaForMaskedLM} encoder-decoder whose parameters are learned from scratch. For domain adaptation to triage, we attach a 4-way softmax classification head (\(\texttt{num\_labels}=4\)) to a RoBERTa encoder initialized from the MLM checkpoint and fine-tune on triage-labeled ED texts from the same train partition. 

To explore the scale-performance trade-off of state-of-the-art LLMs, we fine-tuned three LLMs on the train partition of the dataset: Mistral NeMo 12B \cite{mistral2024nemo}, MedGemma 27B \cite{sellergren2025medgemma}, and gpt-oss-20b \cite{openai2025gptoss20b}.

\vspace{0.2cm}
\noindent\textbf{Training and Inference Procedures.}
Fine-tuning of all LLMs was performed using the quantized model weights loaded through the Unsloth framework for optimized training. Each model was configured with 4-bit quantization using bfloat16 precision to reduce memory requirements while maintaining training stability. 

To enable efficient fine-tuning of the large language model, we used Low-Rank Adaptation (LoRA) as a parameter-efficient fine-tuning approach. The LoRA configuration utilized a rank of 16 with an alpha value of 32 and a dropout rate of 0.1. The adaptation targeted all attention and feed-forward projection layers, as well as the gate, up, and down projections of the transformer architecture.

The training data was formatted using instruction-following templates following the recommended chat format for each model to enable supervised instruction tuning. Each training sample was structured with a system instruction identifying the model as an emergency triage assistant, followed by the patient's clinical information enclosed within clearly marked boundaries, and concluding with the expected triage score response.

For evaluation, test samples were formatted using the same instruction template structure but truncated before the target answer to enable open-ended generation. Inference was performed using a text generation pipeline configured to generate a maximum of 2 new tokens, sufficient to produce only the numerical triage score. The pipeline utilized multi-GPU distribution with load balancing across available GPUs to optimize inference speed. The padding token was set to the end-of-sequence token to ensure consistent tokenization across all samples. Model predictions were extracted from the generated text and parsed as integers corresponding to triage acuity levels ranging from 1 to 4 or 2 to 5, depending on the dataset used.

Three variables from the test portion of the dataset (gender of patient, history of present illness, and past medical history) were selected and assembled into free-text strings.
The \href{https://huggingface.co/mistralai/Mistral-Small-3.2-24B-Instruct-2506}{Mistral Small 24B v3.2 model} was used in a 7-shot learning configuration for automatic gender-swapped pair generation. Seven manually-annotated examples of gender-swapped pairs were included in the model instructions, after the system prompt. The model was instructed to change all references to patient sex/gender to the opposite gender while maintaining all other clinical information unchanged. The model was instructed to respond in comma-separated format (gender of patient, history of present illness, past medical history), allowing storage of transformed data following the same structure as original data. The model, along with its prompt, was served locally using vLLM \cite{kwon2023efficient}.

While the Mistral small model was selected for its balance between size and performance at the time of the execution of experiments, we believe any state-of-the-art reasoning model of similar or greater size could be used for this task.

We frame the task of learning emergency triage from patient data as an ordinal regression problem, where the goal is to train a model that predicts the correct triage category. Each patient record combines two modalities: structured tabular data (vital parameters, demographics, contextual information) and unstructured textual data (patient-reported symptoms, medical history, clinical narratives). The output space consists of ordinal triage scores ranging from 1 (highest urgency) to 5 (lowest urgency). We fine-tune a pretrained LLM to act as a classifier over these triage categories. Each patient record is rendered as a prompt that linearizes the tabular data, appends the text, and instructs the model to output a triage score. The model is trained using standard cross-entropy loss to generate the correct label token as the next token. Formal mathematical definitions of all metrics are provided in Supplementary Note 1.

For generating gender-swapped pairs, we construct modified versions of patient records by flipping the recorded sex in the tabular data and rewriting gendered expressions in the text while preserving all other clinical content. This design is inspired by fairness testing principles that examine how outcomes change when protected attributes are modified \cite{kusner2017counterfactual}. The tabular transformation simply flips the binary sex indicator while holding all other variables fixed. The textual transformation uses a rewriting function that minimally edits explicit and implicit gendered mentions (pronouns, gendered nouns, kinship terms, honorifics) to be consistent with the target sex, implemented via few-shot prompting of an LLM. We apply the transformation only to records for which a gender-consistent version can be produced without changing clinical meaning; records containing inherently gender-specific content that cannot be made consistent without altering medical semantics (e.g., pregnancy-specific content) are excluded from the paired analysis.

We analyze the predictions on each matched pair (the original record and its gender-swapped version) to quantify systematic differences in how the model responds to gender using several complementary metrics. The \textit{Pairwise Disagreement Rate} (PDR) quantifies the proportion of matched pairs for which the model's predicted triage score changes after applying the gender-swap transformation; a high PDR indicates that gender information substantially influences predictions. \textit{Directional Change Probabilities} quantify the direction of change in predicted severity when flipping gender, specifically measuring the probability of upward shifts (toward less severe scores) and downward shifts (toward more severe scores) for each transformation direction (male-to-female and female-to-male). The \textit{Net Mean Difference} (NMD) computes the average signed difference in predicted triage scores between the female and male presentations of the same clinical content; a positive NMD indicates that female presentations receive higher scores (less severe triage) on average, and a negative NMD the opposite. The \textit{Net Asymmetric Triage Shift} (NATS) indices capture directional asymmetries in how predictions change: NATS(+) measures whether upward (less severe) shifts are more common in one transformation direction than the other, while NATS(---) does the same for downward (more severe) shifts. Finally, the \textit{Directional Triage Skew} (DTS) summarizes, within each transformation direction, the net tendency toward higher versus lower predicted severity, measuring the balance between upward and downward shifts when transforming from one gender to the other. Because the two DTS values are computed on disjoint cohorts (the male→female skew on originally-male records and the female→male skew on originally-female records) and each captures the same underlying event from one side, namely a net tendency for the female presentation to be scored less severe than the male presentation of the same content, we also report a single \textit{pooled per-pair rate}: the size-weighted average of the per-cohort female-less-severe rates over all pairs,
\[
\widehat{r}_{\text{pool}} \;=\; \frac{1}{N_{\text{pairs}}}\sum_{i=1}^{N_{\text{pairs}}} e_i,
\qquad
e_i \in \{-1, 0, +1\},
\]
where $e_i=+1$ if the female presentation of pair $i$ receives a higher (less severe) predicted score than the male presentation, $e_i=-1$ if it receives a lower (more severe) score, and $e_i=0$ for ties. This pooled rate is the appropriate single-number summary of the female-versus-male asymmetry; it is not the sum of the two DTS values, which would double-count the same contrast. Confidence intervals for $\widehat{r}_{\text{pool}}$ are obtained by the same nonparametric bootstrap (1{,}000 resamples over pairs) used for the other metrics.

To understand whether bias operates primarily through explicit tabular gender indicators or implicit textual cues, we conduct modality-isolated analyses. In the text-isolated analysis, we remove the tabular sex field and flip only textual gender cues. In the tabular-isolated analysis, we remove the free text and flip only the tabular sex indicator. Comparing metrics across these conditions reveals which modality contributes more to observed disparities.

When we observe asymmetric changes in predicted severity across gender-swapped versions, these effects may originate from (i) the training dataset (e.g., human or data-collection artifacts) or (ii) biases absorbed during general-domain pre-training of the LLM. To disentangle these sources, we compare two model variants: one pre-trained on general-domain text and then fine-tuned for triage, and one trained entirely from scratch on the ED data without general-domain pre-training. If both models show similar bias patterns, this suggests the training data itself contains the bias; additional asymmetry in the pre-trained model would implicate biases acquired during pre-training.

\vspace{0.2cm}
\noindent\textbf{Hardware specifications.}
All fine-tuning and gender-swapped pair generation tasks were executed on a server running Ubuntu 22.04 LTS with an AMD EPYC 7713 Processor and four NVIDIA A100 80GB GPUs.

\vspace{0.2cm}
\noindent\textbf{Regulatory approval.}
This study adhered to the MR-004 reference methodology (research not involving the human person, health studies and assessments for reused data) as outlined by the French National Commission on Informatics and Liberty (CNIL), within the Bordeaux University Hospital. This study and its subsequent publications have received ethics board approval by the Health and Research Ethics Centre of Bordeaux (\textit{Centre Éthique et Recherche en Santé Bordeaux}), with reference CER-BDX 2025-412.

\section*{Data Availability}
Version 2.2 of the MIMIC-IV dataset is publicly available \href{https://physionet.org/content/mimiciv/2.2/}{at this address} upon approval of a user's profile. The Bordeaux University Hospital ED dataset cannot be made publicly available due to current patient privacy regulations in France but may be accessible through formal collaboration agreements.

\section*{Code Availability}
The code for model training, gender-swapped pair generation, and bias analysis is available for download as supplementary material to this article.

\section*{Acknowledgments}
We sincerely thank the triage nurses who participated in this study. While their daily dedication to patient care is a given, it is their contribution to this research that made it possible, and we hope its outcomes will in turn help improve their work. The research efforts of A.G.A. are supported by a doctoral grant from the Digital Public Health Graduate Program of the University of Bordeaux (CD EUR DPH). L.A.C. is funded by the National Institute of Health through DS-I Africa U54 TW012043-01 and Bridge2AI OT2OD032701, the National Science Foundation through ITEST \#2148451, and a grant of the Korea Health Technology R\&D Project through the Korea Health Industry Development Institute (KHIDI), funded by the Ministry of Health \& Welfare, Republic of Korea (grant number: RS-2024-00403047). Data collection and processing is made possible with funding from the TARPON project of the Health Data Hub, France.

\bmhead{Author Contributions}
A.G.A., M.A.F., C.G.J., and E.L. conceptualized the study. A.G.A. performed all experiments, data manipulation, and data analysis, and drafted the original manuscript. O.D. and L.A.C. contributed to methodology development. E.L. and M.A.F. handled project direction, and supervised the research. C.G.J. and L.A.C. provided expertise in emergency medicine. All authors (A.G.A., M.A.F., O.D., L.A.C., C.G.J. and E.L.) reviewed and approved the final version of the manuscript and agree to be accountable for all aspects of the work.

\bmhead{Competing Interests}
The authors declare no competing financial or non-financial interests.

\end{document}